# On Mach's critique of Newton and Copernicus

Herbert I. Hartman
William Rainey Harper College, Palatine, Illinois 60067

Charles Nissim-Sabat
Northeastern Illinois University, Chicago, Illinois 60625,
& Cherskov and Flaynik, Chicago, Illinois 60606

By maintaining the relativity of all motion, especially rotational motion, Mach denied the existence of absolute motion and of absolute space. Accordingly, he maintained the equivalence of the Ptolemaic and the Copernican systems and the equivalence of a rotating bucket in a fixed universe with a fixed bucket in a rotating universe. An analysis of the Foucault pendulum shows that Mach's relativity principle implies that there cannot be a fixed bucket in a rotating universe. Also, Mach's views violate the physics Mach espoused: non-inertial experiments, e.g. stellar aberration, electromagnetic effects, distinguish between a rotating bucket in a fixed universe and a fixed bucket in a rotating universe, between a Copernican universe and a Brahean or Ptolemaic universe and establish that one cannot ascribe all pertinent observations solely to relative motion between a system and the universe.

## I. INTRODUCTION

Physicists and philosophers have debated many topics under the heading of Mach's principle.[1] We will restrict ourselves to Mach's related criticisms of Newton's interpretation of his rotating bucket experiment and of the prevailing view that the Copernican system describes reality. We focus on three statements by Mach that often preface discussions of Mach's principle,[2] and examine the internal consistency of Mach's views and their validity in the light of classical physics, that is, of what Mach knew[3] in 1912, but without using special relativity (SR), which Mach rejected. The extension of our arguments to relativistic physics is obvious. Many of our arguments are independent of each other.

Newton had stated that "The effects which distinguish absolute from relative motion are the [centrifugal] forces. If a vessel … filled with water ... is whirled about [for a while] the surface of the water will be plain; but after that [the water] … will revolve … and … form a concave figure till … it becomes relatively at rest in the vessel … The true and absolute motion of the water, which is here contrary to the relative [motion with respect to the vessel] … may be measured by this endeavor."[4] Here, absolute motion means motion with respect to absolute space.

In the rotating bucket the water is accelerated. Galilean and special relativity are valid only for uniform rectilinear motion. Pre-SR electrodynamics posited that one could measure both velocity and acceleration with respect to a luminiferous ether, and, acceleration with respect to the ether was taken to be identical to acceleration with respect to absolute space. Calling absolute space a "monstrous conception,"[5] Mach rejected the notion of absolute space, but not that of the ether. He claimed that the centrifugal force in the bucket could be due to the water's motion with respect to the masses in the universe. He thus sought to extend to rotational motion the notion that motion is purely relative. We present here a critique of Mach's arguments for the relativity of rotational motion.

**II. ROTATING THE UNIVERSE**

Mach rejected the concept of absolute space and advocated the concept of the relativity of rotation in statements such as: "Try to fix Newton's bucket and rotate the heaven of fixed stars and then prove the absence of centrifugal forces" (Statement 1).[6] We shall assume that Statement 1 was intended to encompass Coriolis forces, although Mach was silent on this point. Statement 1 suggests a Gedankenexperiment, but not a typical Gedankenexperiment, not only because the result is unknown, but also because Mach did not specify the procedure to be followed. Finally, the Gedankenexperiment violates Mach's own precept that in a Gedankenexperiment, "it is permissible to modify [only] unimportant circumstances to bring out new features."[7]

**IIA. How can the bucket be fixed as the universe rotates?**

By "fix" Mach did not mean "fix with respect to absolute space." We assume that for Mach, a fixed bucket is one supported without friction and at rest with respect to the fixed stars before the universe is made to rotate. Whether the universe can be rotated independently from the bucket will be examined below (see Sec. IIF).

**IIB. Mach's Gedankenexperiment and absolute space.**

From here on we shall assume that the universe rotating (UR) experiment can be performed and that a concavity then appears that is identical to the one when the bucket is rotated (BR). A "relationist" (one who denies the existence of absolute space) would conclude from the appearance of the cavity that the concept of absolute space is superfluous. An "absolutist" could argue that the experiment shows that there are two ways to induce centrifugal forces, all within the framework of absolute space.

Consider Einstein's case when he formulated SR: it was known for fifty years that identical eddy currents could be produced two ways, either moving a conductor in the field of a magnet or moving the magnet past the conductor. Yet no one thought that this observation showed that the two ways are inherently the same and that the notion of the ether is superfluous. Certainly not Mach! To prove that the ether is superfluous, Einstein had to propose a principle of relativity (SR) stating that the laws of physics are the same in all inertial frames and to propose experiments testing SR.

Similarly, by arguing for the equivalence of rotating-system/fixed-universe (BR) and universe-rotating/fixed-system (UR) situations, Mach argued for a relativity theory, "Machian relativity," wherein the laws of physics depend only on the relative acceleration between the laboratory and the masses of the universe.[8] Many proponents of SR rejected Machian relativity. Planck criticized Mach in 1910 for the "useless thought that the relativity of all translational motion corresponds to a relativity of all rotary motion."[9] To discover the outcome of the Statement 1 experiment, we must test the predictions of Machian relativity.

In classical physics, the equivalence of the BR and UR scenarios can be tested by means other than inertial forces (Coriolis or centrifugal), and with all observations confined to the laboratory, that is, to the bucket.

(1) An electrically charged liquid in a rotating bucket produces a magnetic field that is determined by the charge density and angular velocity of the liquid. An electric field is induced in a conducting rod that is above the liquid, perpendicular to the bucket axis, and rotating with the bucket. Can one explain these magnetic and electric fields with a rotating (but presumably uncharged) universe?

(2) A rotating charged liquid radiates electromagnetic radiation that carries energy away. So one must do work to keep the bucket rotating at a constant speed. Can one explain the radiation flux and the work done on the bucket in the UR scenario?

(3) If work is not provided, the bucket spin slows down and the kinetic energy lost equals the energy radiated. With UR, the angular velocity of the universe decreases at the same rate as the angular velocity of the bucket with BR, but the kinetic energy lost is much larger.

(4) Sagnac's experiment, first suggested by Michelson in 1904,[10] demonstrated a fringe shift between co-rotating and counter-rotating light beams traveling in a rotating polygon. Observers on the fixed stars may perform the same electro-magnetic and Sagnac experiments ("non-inertial experiments") to determine whether the universe is rotating.

(5) Astronomical observations contradict Machian relativity. Consider stellar aberration in the framework of pre-SR with a luminiferous ether and no limit on the magnitude of the velocities. For a rotating bucket of radius $r$ and angular velocity $\omega$, the relative azimuthal velocity at the conjunction between a star and an observer on the bucket's rim is $\omega r$. The maximum aberration angle for a star as seen from the rim is $\pm \omega r/c$. With UR, the velocity $v$ of a star at a distance $\rho$ from the rotation axis is $\omega \rho$ and $\omega \rho \gg \omega r$. The star's motion through the ether generates a bow wave,[11] with an "optical boom" conical shock wave at an angle $\sin^{-1}(c/v)$ with respect to $v$, if $v > c$. With UR and $\omega > 10$ rpm, the moon has $v > c$, and its light and that of all other bodies propagates as a shock wave. Thus, classical predictions of stellar observations in the BR and UR cases differ qualitatively. (So do predictions under SR.) Hence, Feynman stated: "There is no

'relativity of rotation'." [12]

**IIC. Can one say that the bucket and the universe are merely rotating with respect to each other?**

At first, this relative rotation alternative seems reasonable. After SR, one is tempted to declare it self-evident. But SR does not support this relativity position because the two systems are not inertial. Also, the examples in Sec. IIB argue against the relative rotation hypothesis.

Moreover, one can determine separately the motions of the bucket and of the universe if there is a relative rotation between them. If a bucket displays concavity of the water surface, Newton would say that, from the depth of this concavity, an observer R on the bucket rim can determine $\omega$, the bucket's angular velocity. Given a star S and the bucket center O, the observer R can determine $\omega'$, the relative angular velocity, from the time between successive ORS alignments. For Mach, $\omega = \omega'$ always. For Newton, $\omega = \omega'$ only for BR. A finding by R that $\omega \neq \omega'$ falsifies Machian relativity. Let $V$ be the velocity of R relative to S when O, R, and S are aligned. R can measure $V$ by using stellar aberration. For BR, $V = \omega'r$; for UR, $V = \omega\rho$. If $\omega'r < V < \omega'\rho$, and the bucket has angular velocity $e\omega'$ ($0 < e < 1$) then the universe has angular velocity $(e-1)\omega'$. Given that $V=\omega(er+(e-1)\rho)$, R can determine $e$ and the respective angular velocities even if $\omega = \omega'$.

**IID. Bucket and water interactions in a rotating universe**

The bucket and the water interact differently for BR and UR. For BR, the time it takes for all the water to rotate at the same angular velocity as the bucket depends on the properties of the bucket. Newton probably used a round bucket when he observed the delay between the rotation of the bucket and the concavity of the surface. In a bucket shaped like a thin vertical box, the water will exhibit the maximum concavity immediately after the bucket atarts rotating. With UR, Statement 1 implies that there is only a radial force on the water, so one would expect the concavity to form at the same rate in all buckets. Also, obstacles on the bucket floor and walls would act differently in the UR and BR scenarios.

Also, when the bucket leaks with BR, water droplets fly off tangentially at a constant velocity (neglecting gravity). With UR, one would observe a radial force on the water proportional to the distance from the axis. Would not the same radial force affect the water droplets that have leaked out of the bucket so that they would fly off radially?

**IIE. What if you have several buckets?**

Consider two coaxial buckets A and B rotating with equal and opposite angular velocities. One would observe the same concavity in the two buckets. The extension of Newton's reasoning is straightforward and treats A and B equally and independently. A literal application of Mach's Statement 1 would have the universe rotating in two opposite directions. Presumably, Mach would propose that one posit that A (or B) is fixed, that the universe, bucket B (A) included, revolves around A (B), and that bucket B (A) rotates with respect to the universe. Mach would have to propose two different types of motion for what Newton explains with only one. Mach also would have to invoke even more complex explanations for non-coaxial buckets with arbitrary angular velocities. A and B can be interchanged, but still, one cannot treat the two buckets equally. Mach's account is even more unwieldy when there are many buckets, with non-parallel axes, etc.

The rotating-universe explanation becomes even more strained when one considers gyroscopes. In a gyroscope rotating under gravity with a point fixed and a non-vertical angular momentum, one observes precession and nutation, without the gyroscope falling to the ground. Suppose that we stop the gyroscope spin but rotate the universe. How would Mach predict the same gyroscope motion? Again, one may complicate matters further by considering combinations of gyroscopes, etc.

Newton's absolute space explanation has the advantage of giving the same explanation for what Mach would consider different phenomena. When it comes to rotation, the notion of absolute space functions for Newton the same way as the elimination of the ether does for Einstein by providing a unifying explanation for what would otherwise be an array of disparate and complicated accounts.

**IIF. Machian relativity and the rotation of the universe**

To examine the requirements of Machian relativity, consider rotations and revolutions for half-a-period effected in the plane of the paper , that is, around axes orthogonal to that plane. Also, we shall use objects that are not symmetric around axes orthogonal to that plane (that is they have a "front" and a "back") so that observers can determine relative rotations between these objects. Take a fixed star that has a bright spot. The star is symbolized by S>, with the bright spot designated by the symbol >. A bucket has a V painted on the bottom and is symbolized by V.  One would start the Statement 1 experiment as follows:

        S>           V                                              *Initially*

With BR, as seen from the star, V turns around and becomes Λ, so we have:

        S>           Λ                                          *Bucket Rotating*

With UR but without altering the angular momentum of the star:

                V           S>            *Universe Rotating- Case I*

The fixed star presents a different side to the bucket after half a rotation of the universe, so that the rotation of the universe can be detected by examining visible features of the fixed stars. Thus Machian Relativity requires that as one performs the Statement 1 Gedankenexperiment one must make the fixed stars perform rotations as they revolve around the bucket (or, better yet, have the universe rotate as a rigid body), yielding:

                V           <S            *Universe Rotating-Case II*

Thus Statement 1 does not provide an unambiguous procedure for the Gedankenexperiment.

But is the UR-II scenario the one necessary to produce centrifugal forces in the bucket? It is not obvious that rotation of the fixed stars would have such dynamic consequences on the bucket.  We do not know whether UR-I or UR-II or neither or both will produce centrifugal forces.  Thus the Statement 1 Gedankenexperiment may not advance our understanding of the origin of centrifugal forces.

General Relativity (GR) predicts a Coriolis-type force inside a rotating shell and this has inspired many to invoke the Foucault pendulum as a prime instance where Mach's approach must be applied. Mach gave the Foucault pendulum but scant attention. A Gedankenexperiment with a Foucault pendulum yields surprising results.

Consider a pendulum freely mounted above the bucket and swinging with the same frequency as the bucket's rotation. Let S>, Δ, and V denote the positions of a fixed star, the pendulum bob, and the bucket. The pendulum bob has a mark such as Δ, while V and S> remain as above. Initially, with the pendulum fully extended, we have:

 S>    Δ  V           *Initially*

Half a period later, and with BR, we have:

 S>      Λ  Δ        *Bucket Rotating*

With the same initial conditions, half a period later, but with UR-II:

      V  Δ      <S  *Universe Rotating-II*.

The two scenarios are distinguishable in that with UR the bucket never finds itself between the pendulum bob and the fixed star. (The bob is directly above the bucket at time = 1/4 or 3/4 period). One may object to our UR configuration because we have the pendulum swing aligned with respect to the bucket rather than the fixed star. We postulate that the rotation of the universe induces a precession of the pendulum with an angular velocity that is exactly equal to the angular velocity of UR. Adopting this model, the UR configuration becomes

      Δ……V      <S  *U. Rotating but Δ tracks  S>*.

Note that Δ and V have their markings aligned contrary to the BR case. For the BR and UR alignments to be the same one must postulate that rotation of the universe induces a rotation in the pendulum bob, so that Δ becomes ∇.[13] But if a rotation is induced in the pendulum bob, should not a rotation be induced in the bucket as well, yielding:

      ∇  Λ       <S  *UR, all objects tracking S>*,

But now, after the bucket's rotation and the pendulum's swing, the pendulum, bucket, and /universe relative alignments are identical to the initial configuration and different from the BR configuration. But, the bucket's rotation is independent of whether there is a pendulum swinging overhead and the pendulum bob could have been a bucket. Moreover, the bucket's markings are immaterial to its dynamical properties. Therefore, Machian relativity requires that the rotation of the universe induce an equal rotation of a freely mounted bucket at the center of the universe. This of the bucket under UR may be said to be obvious under Machian relativity: if a bucket starts out in alignment with a given star, it would have to remain aligned with that star under UR for there is no other reference point with which it can remain aligned, certainly no point in absolute space because Mach denied the existence of absolute space.

That Machian relativity requires that rotation of the universe induce rotation of a freely mounted bucket at the center of the universe also can be seen as follows. Consider two coaxial buckets A and B, one of which (we do not know which) will be made to rotate by some random process. When one of the buckets (say A) rotates in a fixed universe, it will exhibit centrifugal forces while B will remain aligned with a fixed star and exhibit no centrifugal forces. But let us do the UR experiment first. Which bucket will remain aligned with the fixed star? Nothing distinguishes the two buckets but a future choice. So either both buckets rotate and exhibit centrifugal forces as the universe rotates, or neither one does, regardless of which bucket will rotate later with respect to a fixed universe. Thus Machian relativity requires that all objects at the center of a rotating universe rotate, even if they are not mechanically coupled to the universe. This requirement also is true at the microscopic level: $\Delta$, V, and <S could represent electron spins.

Thus, the theory of relativity for rotational motion that Mach based on relationism leads to a fundamental contradiction in that there cannot be a non-rotating bucket at the center of a rotating universe, even if bucket and universe are mechanically decoupled. Also, astronomical, optical, and electromagnetic effects distinguish between rotating-laboratory/fixed-universe, rotating-universe/fixed-laboratory, or admixtures of these situations. Einstein stated in 1934: "I was ... acquainted with Mach's view that ... acceleration [is] but acceleration with respect to the masses [of the universe]. There was something fascinating about this idea, but it provided no workable basis for a new

theory."[14]

**III. IS THE PTOLEMAIC SYSTEM ACTUAL?**

An analysis of the consequences of Statement 1, that is, of Machian relativity, can, as Mach himself insisted, be carried out by comparing the heliocentric and geocentric systems: "The system of the world is given ***once*** to us, and the Ptolemaic or Copernican view is our interpretation, but both are equally actual …The motions of the universe are the same whether we adopt the Ptolemaic or the Copernican mode of view. Both are indeed equally ***correct***; only the latter is more simple and more ***practical***. The universe is not **twice** given, with an earth at rest and an earth in motion; but only **once**, with its **relative** motions alone determinable" (emphasis in the original). (Statement 2).[15]

In Statement 1 it is the bucket, and not the universe, that is actually rotating. Rotating the universe is a Gedankenexperiment. In Statement 2, the two possibilities, (a) the earth moving/the universe fixed, and (b) the earth fixed/the universe moving, are equally actual. Perhaps the difference between the statements stems from the human agency in the bucket's rotation, while we are given the heavens' apparent motions.

We distinguish (but Mach did not) between the heavens' (1) daily and (2) annual motions, and between the (2a) Ptolemaic system and (2b) Brahe's version thereof, where all the planets (except the earth) orbit the sun and the sun orbits the earth. The questions raised in Sec. II apply to the Copernican, the Ptolemaic, and the Brahean systems. Using the non-inertial experiments mentioned in Sec. IIB, could not earth-bound, sun-bound, and fixed star observers detect their own motions and determine their orbits, including whether the sun is revolving around the earth?

**IIIA. Is the universe rotating with a 24-hour period?**

Mach did not address a daily rotation of the universe explicitly. He stated that the Foucault pendulum proves that the earth rotates.[16] Yet the Ptolemaic/Brahean system had the universe rotate daily, and the "equally actual" portion of Statement 2 immediately precedes Statement 1, so we conclude Statement 2 implies a daily rotation of the

PAGE 10

universe.

Some considerations from Sec. II specifically apply here. Can there be a fixed earth at the center of a rotating universe? Also, were the sun and the stars to orbit the earth in 24 hours, we would see 24-hour stellar aberration effects which, because of the much larger velocity of the fixed stars, far exceed those presently observed. Even $\alpha$ Centauri is far enough to produce the "optical boom" discussed in Sec. IIB. This optical boom would enhance the luminosities of distant stars and of stars on the earth's equatorial plane and yield a distorted and anisotropic stellar distance/luminosity relation.

Nor can Mach explain the many effects Newtonian physics ascribes to the earth's rotation. Consider only the slowing down of the diurnal period (from 12 h to 24 h over a billion years) due to the friction of the oceans on the ocean floors because of the lunar tides. This slowing down corresponds to a loss of 3/4 of the earth's rotational kinetic energy over that time. Half of the earth's rotational angular momentum is lost at the same time: it is converted into the orbital angular momentum of the moon as it recedes from the earth. Can Mach say that the earth has been standing still (not rotating) all along, and it is the universe's rotation that has slowed down by a factor of two? The universe's moment of inertia is much larger, and we are left to account for enormous energy and angular momentum losses.

Also, had the earth actually been still at any one time, the torque produced by the moon's gravitational interaction with the tidal bulges would have produced rotation of the earth. Thus, unless the earth is rigidly held in place, the earth is now rotating.

**IIIB. What did Mach mean by Ptolemaic?**

Concerning the heavens' annual motion, the Ptolemaic system (2a) was falsified by Galileo's observation of the full disk of Venus, possible only when Venus is on the opposite side of the Sun from the earth. Mach had stated that Galileo's discoveries were "new and very strong arguments for Copernicus."[17] Why was Mach not convinced? Did Mach have Brahe's (2b) system in mind: the sun orbits the earth while all the planets but the earth orbit the sun? The observation of the full disk of Venus does not falsify Brahe's system. Galileo made a dead horse of the Ptolemaic system, so we will address specifically Brahe's system, but our remarks apply to the original Ptolemaic system as well.

Brahe formulated his system before it was established that the planets' orbits, including the earth's, are elliptical with the center of mass of the solar system at the focus, with perturbations of the orbits due to their interactions with their respective satellites and with each other, and, we would add today, with the precessions that Einstein explained. Stellar aberration, parallaxes, and Doppler shifts allow us to measure the earth's motion relative to the stars. In view of these properties of the orbits, a Brahean system has the earth fixed, the other planets orbiting the solar system center of mass and the center of mass orbiting the earth in a most irregular annual orbit. In the approximation where the "fixed stars" are at rest with respect to the solar center of mass, the fixed stars would duplicate the same irregular orbit as the sun in orbits parallel-displaced to the sun's orbit. In Brahe's system, there is no mass at a focus of a stellar orbit. Mach ignored these complications that make Brahe's system most implausible.

**IIIC. Are both systems equally correct, equally actual?**

With his equally correct assertion, was Mach merely echoing the famous but unauthorized preface that Osiander added anonymously to Copernicus' *De Revolutionibus?*[18] That was in 1543, but by 1912 the disk of Venus, the satellites of the planets, the Foucault pendulum, and stellar aberration, parallaxes, and Doppler shifts had all been observed.

It is a truism that one can describe any phenomenon from any reference point. An observer on earth would say that an apple falls to the ground with an acceleration of 9.8 m/s$^2$. An apple-bound observer would say that the earth rushes to the apple with an acceleration of 9.8 m/s$^2$. Also, Brehme[19] has shown that one can obtain straightforwardly a geocentric system from the Copernican system. Yet, the equivalence in these examples is purely kinematical. However, Mach would insist that kinematic equivalence is all that counts: kinematics records observations, that is, facts, and Mach maintained that science should adhere to "actual facts"[20] focusing on how questions, not why questions.[21] But, science so construed would not allow for predictions: facts are all in the past. Also, Mach neglected that there are no theory-free facts and provided no reason why luno-, jovo-, venocentric, etc., systems are not as actual as the geo- and heliocentric ones.

Bunge studied Mach's approach to mechanics and concluded that "the dynamics of a mechanical system [cannot be] inferred from observations of its parts ... [As] Newton had discovered, kinematics is deducible from dynamics but not, as ... Mach wanted, conversely." Dynamics, the laws of nature, need "non-observational" (that is, theoretical) concepts for their discovery.[22] Similarly Einstein stated: "Mach's epistemological position influenced me very greatly, a position which today appears essentially untenable. He did not place in the correct light the essentially constructive and speculative nature of thought, more especially, of scientific thought."[23] But, it turns out, kinematical equivalence is not always true. As Kant first noted, the irreducible difference between the right and left hands argues against a relational theory of space.[24] Observers using a mirror may make different observations. Given parity non-conservation, this left/right asymmetry is true even at the microscopic level: neutrinos are left-handed in this world, right-handed in a mirror, and, in the mirror, they behave differently from the right-handed neutrinos (antineutrinos) in this world.

There is no indication that Mach intended a principle of general covariance (GC). Einstein[25] did indicate that Mach had influenced him in formulating that principle, but, for Einstein, GC for the laws of physics was an outgrowth of Lorentz invariance and required that the laws be formulated in tensor form in a four-dimensional spacetime. Mach rejected Lorentz invariance and did not attempt to formulate a generally covariant theory. GC does not support Statement 2. GC holds that the laws of physics must be formulated in the same tensor form in all reference frames, but inertial frames can be distinguished from all others: the laws of SR physics apply in inertial frames and observers can determine whether they are in such a frame. In the falling apple case, an earth-bound observer would determine that, with gravitation, $F = ma$ for the apple, but an apple-bound observer would deny that $F = ma$ for the earth.

By insisting that the sun defines an inertial frame (or nearly so) while the earth does not, Bunge dismisses Statement 2 as a "grotesque claim" not withstanding that it is advanced by many authorities and "hundreds of textbook authors."[26] The laws of physics do apply in a heliocentric system but not in a geocentric (Brahean or Ptolemaic) system. When Ptolemy and Brahe formulated their systems, they knew of no law of physics that their systems would violate. Mach knew that the geocentric systems violated the laws of physics, but he never addressed how the heliocentric and the geometric systems can be considered equally actual if the laws of physics apply in one system but not in the other. Nor did Mach formulate an alternative set of laws to explain the motions of the stars, the planets, the sun, and the moon in a Brahean system.

**IIID. Are there conservation laws with Mach?**

The sterility of a physics without laws also can be illustrated by considering what happens when one has no recourse to conservation laws. For Newton, given bodies A and B, and U, the rest of the universe, **p**A + **p**B + **p**U = **p'**A + **p'**B + **p'**U for the momentum before and after an event. If **p**A+**p**B ≠ **p'**A +**p'**B, then the experimenter is led to seek a particle or field that transfers momentum from the (A, B) pair to the universe. For Mach **p**A + **p**B + **p**U is not defined and no conclusions can be drawn when **p**A + **p**B differs from **p'**A + **p'**B. The same analysis applies to the laws of conservation of energy and of angular momentum.

A naive Copernican account of the earth's revolution around a fixed sun is not actual. What is meant today by 'the Copernican system' is that the sun and all the planets orbit around their common center of mass. The Brahean system violates conservation of momentum in that the solar system does not orbit around the center of mass and Mach gives no inkling on how to deal with this.

**IIIE. The classical Doppler formulae do not support Mach**

The classical Doppler effect provides crucial observations that relate to the equal actuality of the Brahean and Copernican systems. In the two systems, the relative velocity of the earth with respect to a light source is the same. Yet, classical physics, but not SR, predicts different Doppler shifts for the source moving versus the observer moving. Thus, in classical physics, one could determine whether the earth moves or a "fixed star" moves. This different treatment for the two motions should have motivated Mach to pay closer attention to SR.

To conclude, Mach did not consider the differences between the Copernican and Ptolemaic/Brahean systems and the many observations falsifying the latter. Also, by arguing that the two systems are equally actual, he denied the universality of the laws of physics.

## IV. THE CASE OF THE THICK BUCKET

Mach reproached Newton for not considering the possibility that concavity in the rotating water can be caused by large nearby masses: "No one is competent to say how the experiment would turn out if the sides of the bucket increased in thickness and mass until they were ultimately several leagues thick."[27] (Statement 3).

But, the bucket's thickness and mass may not affect the shape of the rotating liquid. Its shape is determined by the components along the surface of the force of gravity (proportional to the gravitational mass) and of the centrifugal force (proportional to the inertial mass). According to Mach, a thick bucket would increase the mass density of the water (Mach did not differentiate between inertial and gravitational mass). But, with the inertial and gravitational masses being equal, the shape of the surface of the liquid is independent of the density of the liquid.[28]

We do not believe that a theory in which the inertial mass of an object depends on the influence of other masses fulfills per se Mach's strictures against the notion of absolute space. All such a theory can do is alter the $m$ in $F = ma$, $m$ perhaps decreasing as the Universe expands, while leaving $a$ defined with respect to absolute space.

Presumably, Mach believed a thick bucket would cause the water to swirl (and the depression to appear) sooner after the bucket is set in motion. But consider a bucket so massive as to comprise the whole universe, except for a little water at the center. When

PAGE 15

the water's angular velocity equals the bucket's, the water is at rest with respect to the universe and, according to Mach, there would be no depression of the surface. Yet the non-inertial experiments discussed in Sec. IIB would detect the bucket's rotation. Conversely, consider a very thin rotating bucket containing a large amount of co-rotating liquid, in the limit where the liquid comprises the whole universe except for the bucket. Mach would argue that the centrifugal force in the liquid would vanish because there is no mass that can cause the force or with respect to which one can detect rotation. Yet, again, the non-inertial experiments would detect rotation.

## V. CONCLUSION

Mach attempted to disprove the existence of absolute space by maintaining the equivalence of a rotating bucket in a fixed universe, and a rotating universe with a fixed bucket and, more generally, by positing a principle of relativity for rotational motion. We have shown that this relativity principle leads to the contradiction that there cannot be a fixed bucket in a rotating universe. In addition, Mach's views require a thoroughgoing revision of the laws of classical physics, including laws to which he subscribed. Yet, there are experiments, stellar aberration among them, that distinguish between a rotating universe and a fixed one, between a heliocentric universe and a geocentric universe. These experiments preclude an account of these experiments and observations solely in terms of relative rotations.

We are puzzled by the failure of Mach and others to address the questions that we have raised. Perhaps because of Einstein's many statements acknowledging his debt to Mach, some have maintained that general relativity has given us a Machian theory of the universe[29] or will soon do so. General relativity has not done and cannot do so. For instance, the solution to the Einstein field equations that Gödel discovered exhibits distinct anti-Machian features.[30] Einstein repudiated Mach's principle in the end.[31]

Recent experiments have put a damper on Machian hopes. It seems that $dG/dt = 0$ ($G$ being the universal constant of gravitation), that the scalar component that Brans and Dicke added to general relativity to produce Machian effects vanishes, and that there is no effect on the inertia of an object due to the action of nearby masses.[32]

We have discussed the internal consistency and the validity of Mach's views in the light of pre-relativity physics. We have not examined Newton's views in detail nor developments concerning the concepts of space and time in the light of 20th century physics. We refer the reader to Earman's book and to the many works cited therein.[33]

Acknowledgments

We thank an anonymous referee for directing our attention to Bunge's work.